\documentclass[12pt]{iopart}
\usepackage{iopams}
\usepackage[english]{babel}
\usepackage{graphicx}
\usepackage{cite}
\usepackage{bbm}

\begin{document}

\title{The Pegg-Barnett phase operator and the discrete Fourier transform}

\author{Armando Perez-Leija$^1$, L.A. Andrade-Morales$^2$, Francisco Soto-Eguibar$^2$, Alexander Szameit$^1$ and  H\'ector M. Moya-Cessa$^2$}
\address{$^1$Institute of Applied Physics, Abbe School of Photonics, Friedrich-Schiller-Universit\"at Jena, Max-Wien Platz 1,
07743 Jena, Germany
\\ $^2$Instituto Nacional de Astrof\'{i}sica, \'{O}ptica y Electr\'{o}nica \\ Calle Luis Enrique Erro No. 1, Sta. Ma. Tonantzintla, Pue. CP 72840, Mexico }

%%%%%%%%%%%%%%%%%%%%%%%%%%%%%%%%%%%%%%%%%%%%%%%%%%%%%%%%%%%%%%%%%%%%%%%%
% Abstract
%%%%%%%%%%%%%%%%%%%%%%%%%%%%%%%%%%%%%%%%%%%%%%%%%%%%%%%%%%%%%%%%%%%
%
\begin{abstract}
In quantum mechanics the position and momentum operators are related to each other via the Fourier transform. In the same way, here we show that the so-called Pegg-Barnett phase operator can be obtained by the application of the discrete Fourier transform to the number operator defined in a finite-dimensional Hilbert space. Furthermore, we show that the structure of the London-Susskind-Glogower phase operator, whose natural logarithm give rise the Pegg-Barnett phase operator, is contained into the Hamiltonian of circular waveguide arrays. Our results may find applications in the development of new finite-dimensional photonic systems with interesting phase-dependent properties.
\end{abstract}
%%%%%%%%%%%%%%%%%%%%%%%%%%%%%%%%%%%%%%%%%%%%%%%%%%%%%%%%%%%%%%%%%%%%%%%%%%%%%%%%%%%%%%%%

%\ocis{ (050.5298) Photonic crystals;  (230.7370) Waveguides;  (350.5500) Propagation.}
%\submitto{\PS}

\maketitle

%%%%%%%%%%%%%%%%%%%%%%%%%%%%%%%%%%%%%%%%%%%%%%%%%%%%%%%%%%%%%%%%%%%%%%%%%%%%%%%%%%%%%%%%

\section{Introduction}

In classical physics, a wave is described by well-defined amplitude and phase, and both quantities can be measured simultaneously with arbitrary accuracy. In the quantum realm the situation is completely different and up to date a well-defined Hermitian phase operator has been elusive\cite{Dirac,London,Susskind,Louissel,Carruthers,Nieto,Turski,Levy,Loudon,Pegg2,Pegg,Vaccaro3,Bialynicki,Schleich,Perinova,Lynch,Vaccaro,Moya,Royer}.
The lack of such a phase operator has yield many scientists to propose diverse approaches to describe quantum phase. In this regard, one may mention the London-Susskind-Glogower phase operators, the variants of phase space representations, ranking from number-phase Wigner functions to radially integrated quasiprobability distributions \cite{Garraway,Otro,Vaccaro2}. Perhaps the most prominent approach is the so-called Pegg-Barnett (PB) phase operator\cite{Pegg,Barnett}. This PB phase operator is defined in a finite-dimensional state space and due to this finiteness one can define phase states in a rigorous way. The success of the PB formalism lies on the fact that all expectation values of physical variables in a finite-dimensional Hilbert space give real numbers which depend parametrically on the dimension of the state space \cite{Gantsog, Varro}. Certainly, many realistic physical systems are discrete and finite, as a result, the PB theory can be readily used to investigate phase-dependent quantum systems.
\\
Along those lines, optics occupies a notable place since discrete optical systems can be  created either in free space or on-chip, e.g. waveguide lattices \cite{abo}. In fact, over the past decade, the tremendous progress in fabrication and characterization of photonic structures has allowed us to create arrays of evanescently coupled waveguides with a number of channels ranging from a few to a few hundred \cite{Christodoulides,Szameit}. The resulting discrete diffraction makes such coupled configurations a perfect paradigm for the realization of quantum particle tunneling in one or two dimensional lattices, and it permits the observation of quantum and condensed matter phenomena in macroscopic integrated systems using classical and quantum light \cite{Leija,Longhi,Weimann,Grafe,Leija2}. In the waveguides, the local refractive index and the width of the channels determine the on-site potentials (propagation constants), while the tunneling amplitude (coupling coefficients) from site to site is adjusted by changing the separation distance between adjacent waveguides \cite{Szameit2}.
\\
The aim of the present work is to show that the so-called Pegg-Barnett phase operator can be obtained by the application of the discrete Fourier transform to the number operator defined in a finite-dimensional Hilbert space. Then, we elucidate that the structure of the London-Susskind-Glogower phase operator is contained into the Hamiltonian of circular waveguide arrays. As a result, light dynamics arrising in such waveguide configurations will feature properties akin to finite phase operators.

\section{Fourier transform and the phase operator}

In quantum mechanics the position operator ${x}$ and momentum operator $p$ are canonically conjugated variables and they are related via the Fourier operator \cite{Agarwal,Fan}
\begin{eqnarray}
{p} = i^{\hat{n}}x(-i)^{\hat{n}}.
\end{eqnarray}
Here ${\hat{n}}=a^{\dagger}a$ is the number operator, with $a$ and  $a^{\dagger}$ being the annihilation and creation operators, respectively. In terms of $a$ and  $a^{\dagger}$, $x$ is written as
\begin{eqnarray}
x=\frac{a+a^{\dagger}}{\sqrt{2}}.
\end{eqnarray}
In similar manner, $p$ becomes
\begin{eqnarray}\label{PtoQ}
{p} = e^{i\frac{\hat{n}\pi}{2}}\frac{a+a^{\dagger}}{\sqrt{2}}e^{-i\frac{\hat{n}\pi}{2}}=-i\frac{a-a^{\dagger}}{\sqrt{2}}.
\end{eqnarray}
\\
According to Eq.(\ref{PtoQ}), the momentum operator is obtained from the position operator through a Fourier transform.
\\
Following these ideas, we now define an operator $\Phi_s$ obtained by the application of the $(s+1)$-dimensional DFT to the finite number operator $N_s$
\begin{eqnarray} \label{Pegg}
\Phi_s  \propto {\cal F}_s N_s{\cal F}_s^{-1}.
\end{eqnarray}
Here ${\cal F}_s$ represents the DFT, ${\cal F}_s^{-1}$ its inverse, and the number operator $N_s$ is defined as
\begin{eqnarray}
 N_s=\sum_{k=0}^s k|k\rangle\langle k|.
\end{eqnarray}
In matrix form the number operator  is written as
\begin{equation}
    {N_s}=  \left(
        \begin{array}{ccccc}
              0 & 0 & 0& \cdots & 0 \\
              0 & 1 & 0&\cdots & 0 \\
              0 & 0 & 2&\cdots & 0\\
              \cdots&&&&\vdots\\
             0& 0 & \cdots & 0& s\\
        \end{array}
    \right).
\end{equation}
The DFT is given by the Vandermonde matrix \cite{Soto}
\begin{equation}
    {\cal F}_s= \frac{1}{\sqrt{s+1}} \left(
        \begin{array}{cccc}
              1 & 1 & \cdots & 1 \\
              \lambda_0 & \lambda_1 & \cdots & \lambda_s \\
              \lambda_0^2 & \lambda_1^2 & \cdots & \lambda_s^2\\
              \cdots\\
              \lambda_0^{s} & \lambda_1^{s} & \cdots & \lambda_s^{s}\\
        \end{array}
    \right)
\end{equation}
with
\begin{equation}
\lambda_j = \exp \left[  i \frac{2\pi}{s+1} j  \right],  \qquad j=0,1,2,\cdots,s. \label{eigen}
\end{equation}
Note that, because ${\cal F}_s^\dagger={\cal F}_s^{-1}$ the phase operator defined in Eq.(\ref{Pegg}) is Hermitian. In Dirac notation the DFT operator reads
\begin{eqnarray}
{\cal F}_s = \frac{1}{\sqrt{s+1}}\sum_{n=0}^s \sum_{k=0}^s  |n\rangle\langle k|e^{  i \frac{2\pi}{s+1} nk } .
\end{eqnarray}
From the above equations one can readily see that the DFT of the number operator is
\begin{eqnarray}
{\cal F}_s N_s {\cal F}_s^{\dagger}= \frac{1}{{s+1}}\sum_{n=0}^s \sum_{k=0}^s  |n\rangle\langle k| \sum_{m=0}^s me^{  i \frac{2\pi}{s+1} (n-k)m }. \label{transformada}
\end{eqnarray}
By defining the phase states
\begin{equation}
|\theta\rangle =\frac{1}{\sqrt{1+s}}\sum_{n=0}^se^{in\theta}|n\rangle,
\end{equation}
we can rewrite Eq.(\ref{transformada}) in a more compact form
\begin{eqnarray}
{\cal F}_s N_s {\cal F}_s^{\dagger}=  \sum_{m=0}^s m|\theta_m\rangle\langle \theta_m|,
\end{eqnarray}
with $\theta_m=\frac{2m\pi}{s+1}$. Normalization of Eq.(\ref{transformada}) yields \cite{Pegg2}
\begin{eqnarray}\label{PO}
{\cal F}_s \frac{2\pi N_s}{s+1} {\cal F}_s^{\dagger}=  \sum_{m=0}^s \theta_m|\theta_m\rangle\langle \theta_m|,
\end{eqnarray}
which is the so-called Pegg-Barnett phase operator with phase reference set to zero. This demonstrates that indeed the PB phase operator arises as the DFT of the finite number operator.
\\
We can further work on Eq.(\ref{transformada}) to find a simpler form of the transformation
\begin{eqnarray}
{\cal F}_s N_s {\cal F}_s^{\dagger}= \frac{1}{{s+1}}\sum_{n=0}^s   |n\rangle\langle n| \sum_{m=0}^s m+\frac{1}{{s+1}}\sum_{n\ne k}^s   |n\rangle\langle k| \sum_{m=0}^s me^{  i \frac{2\pi}{s+1} (n-k)m }
\end{eqnarray}
which gives
\begin{eqnarray}
{\cal F}_s N_s {\cal F}_s^{\dagger}= \frac{s}{{2}}+\frac{1}{{s+1}}\sum_{n\ne k}^s   |n\rangle\langle k| \sum_{m=0}^s me^{  i \frac{2\pi}{s+1} (n-k)m }.
\end{eqnarray}
Note that the last sum can be cast in a closed form
\begin{eqnarray}
\sum_{m=0}^s me^{  i \frac{2\pi}{s+1} (n-k)m }=-i\frac{s+1}{2\sin(\frac{\pi}{s+1}[n-k])}e^{i\frac{2s+1}{s+1}\pi(n-k)}.
\end{eqnarray}
Thus, we obtain
\begin{eqnarray}
{\cal F}_s N_s {\cal F}_s^{\dagger}= \frac{s}{{2}}+-\frac{i}{2}\sum_{n\ne k}^s  \frac{|n\rangle\langle k|e^{i\frac{2s+1}{s+1}\pi(n-k)} }{\sin(\frac{\pi}{s+1}[n-k])}.
\end{eqnarray}
After some algebra we finally find
\begin{eqnarray}\label{CF}
{\cal F}_s N_s {\cal F}_s^{\dagger}= \frac{s}{{2}}+(s+1)\sum_{n\ne k}^s  \frac{|n\rangle\langle k|}{e^{i\frac{2\pi}{s+1}[n-k]}-1} .
\end{eqnarray}

\section{Circular waveguide arrays and the London-Susskind-Glogower phase operator}

\begin{figure}[h!]
\begin{center}
\includegraphics[scale=0.3]{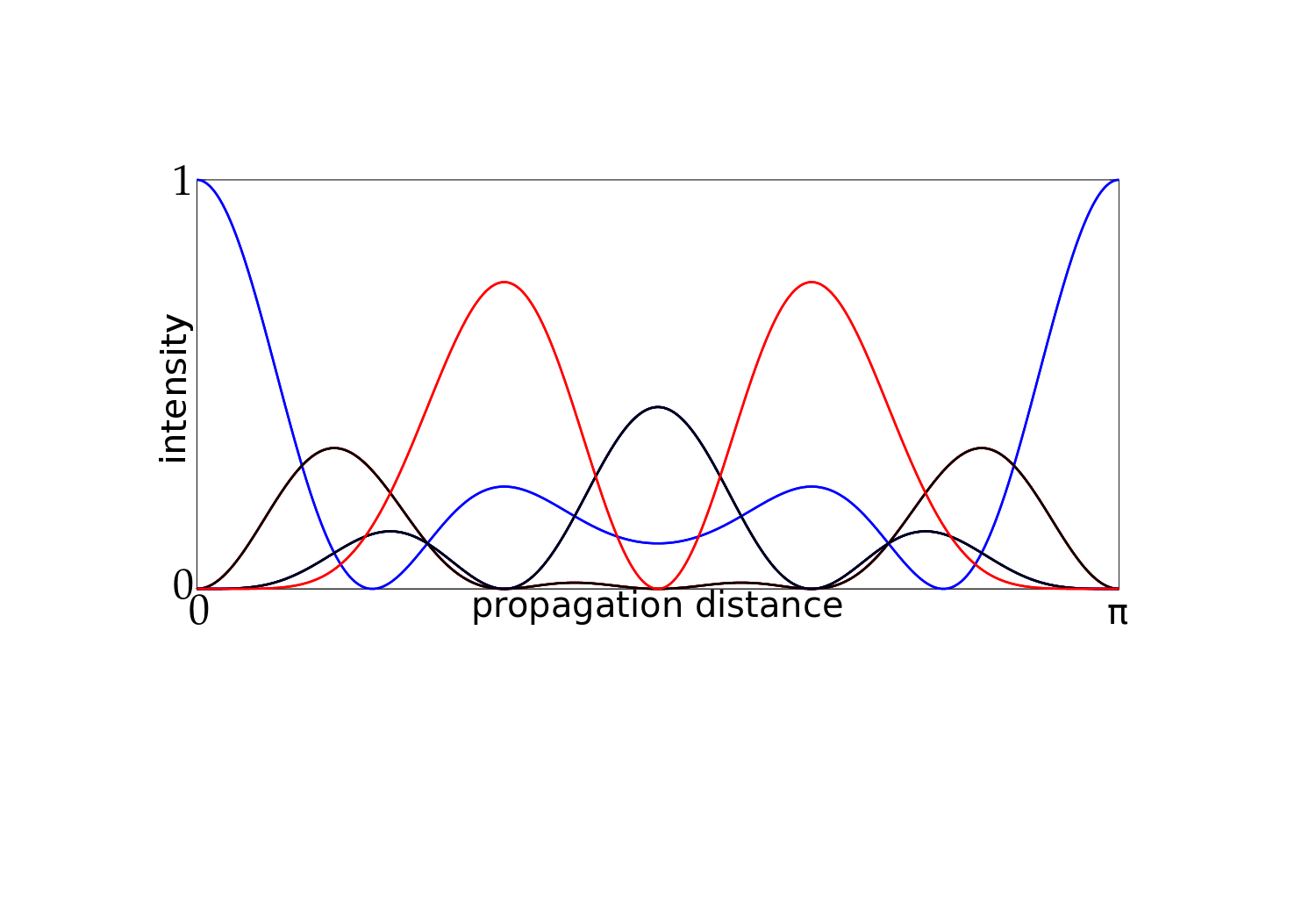}
\caption{Intensity light evolution in a circular waveguide array having six sites. In this simulation we assume a circular array of identical waveguides and equal coupling coefficients $\gamma=1$. The blue curve (online only) depicts the intensity evolution along th excited site. Note that the intensities are periodic and that at $z=\pi$ the intensity of the excited channel becomes one.}\label{fig1}
\end{center}
\end{figure}

In this section we show that the Hamiltonian describing light propagation in circular waveguide arrays explicitly contains the structure of the London-Susskind-Glogower (exponential) phase operator. To do so, we start by considering an array of waveguides in a circular configuration and let the individual guides interact only to first neighbours, the field in each waveguide obeys  the following system of differential of equations
\begin{eqnarray}\label{sistema1}
        i \frac{d E_0}{dz}&=\gamma \left( E_s + E_1\right)\nonumber \\
        i \frac{d E_{n}}{dz}&=\gamma \left( E_{n-1} + E_{n+1}\right), \qquad n=1, ..., s-1.\\
        i \frac{d E_s}{dz}&=\gamma \left( E_{s-1} + E_1\right)\nonumber
\end{eqnarray}
Here, $z$ represents the propagation distance and $\gamma$ the coupling constants.
By introducing the matrix
\begin{equation} \label{London1}
    {V_s}=  \left(
        \begin{array}{cccccc}
              0 & 1 & 0& \cdots & 0&0 \\
              0 & 0 & 1&0 &\cdots & 0 \\
              0 & 0 & 0&\cdots &1& \vdots\\
              0&0&\cdots&0&0&1\\
             1& 0 & \cdots & 0& 0&0\\
        \end{array}
    \right),
\end{equation}
we can cast Eq.~(\ref{sistema1}) as follows
\begin{equation}\label{sistema}
    i \frac{d \vec{E}}{dz}=\gamma \left(V_s+V_s^\dagger\right)\vec{E} ,
\end{equation}
where $V_s^\dagger$ represents the transpose of $V_{s}$, and
\begin{equation}
    \vec{E}=\left(
        \begin{array}{lll}
              E_{0} \\
              E_{2} \\
              \vdots \\
              E_{s}
        \end{array}
    \right).
\end{equation}
Therefore, the field amplitudes over the entire array is given by the formal solution
\begin{equation}
    \vec{E}(z)=\exp\left[-i \gamma z \left( V_s + V_s^\dagger \right)  \right] \vec{E}(0),
\end{equation}
where $\vec{E}(0)$ represents the initial optical field.
Since $V_s$ and $V_s^\dagger$ commute and since they are the inverse of each other, we can write the above solution as
\begin{equation}
    \vec{E}(z)=\exp\left[ -\gamma z \left( iV_s - \frac{1}{iV_s} \right)  \right] \vec{E}(0),
\end{equation}
Hence, by using the generating function of the Bessel functions of the first kind
\begin{equation}
    \exp\left[\frac{x}{2}\left(t-\frac{1}{t}\right)\right]=\sum_{n=-\infty}^{\infty}t^{n}J_{n}(x),
\end{equation}
one can show that the field amplitude is given by
\begin{equation}\label{field1}
    \vec{E}(z)=\sum_{n=-\infty}^{\infty} i^n J_n (-2\gamma z) (V_s)^n  \vec{E}(0).
\end{equation}
Before discussing the properties of this solution, we note that by using Dirac notation $V_{s}$ acquires the form
\begin{equation}
    {V_s}=  \sum_{n=0}^{s-1} |n\rangle\langle n+1| +|s \rangle\langle 0|,
\end{equation}
which is the London-Susskind-Glogower phase operator defined in a $s+1$-dimensional Hilbert space.\\
In what follows we show the functional relationship between the matrix $V_s$ and the PB phase operator.To do so, we consider the eigenvalues of $V_s$, given by Eq.(\ref{eigen}), which allow us to compute any function of $V_s$ through the corresponding Vandermonde matrices\footnote{In this case the similiarity (transformation) matrix and the Vandermonde matrix are the same.}
\begin{eqnarray}
f(V_s) ={\cal F}_s f(D_s) {\cal F}_s^{\dagger} \label{function},
\end{eqnarray}
where $D_s$ is a diagonal matrix having the eigenvalues of $V_s$ as elements
\begin{equation}
    {D_s}=  \left(
        \begin{array}{ccccc}
              1 & 0 & 0& \cdots & 0 \\
              0 &  e^{i \frac{2\pi}{s+1} }  & 0&\cdots & 0 \\
              0 & 0 & e^{i \frac{4\pi}{s+1} }&\cdots & 0\\
              \cdots&&&&\vdots\\
             0& 0 & \cdots & 0& e^{i \frac{2s\pi}{s+1} }\\
        \end{array}
    \right),
\end{equation}
or in Dirac notation
\begin{eqnarray}
 D_s=  \sum_{n=0}^s \exp \left[  i \frac{2\pi}{s+1} n  \right]  |n\rangle\langle n|.
\end{eqnarray}
From Eq.(\ref{function}) is clear that
\begin{eqnarray}
\ln(V_s) ={\cal F}_s \sum_{n=0}^s \ln \left(\exp \left[  i \frac{2\pi}{s+1} n  \right] \right) |n\rangle\langle n| {\cal F}_s^{\dagger},
\end{eqnarray}
or equivalently
\begin{eqnarray}\label{Phi}
\ln(V_s) =i\frac{2\pi}{s+1}{\cal F}_s N_s {\cal F}_s^{\dagger}.
\end{eqnarray}
We note that the right hand side of Eq.~(\ref{Phi}) is the PB phase operator, see Eq.~(\ref{PO}). As a result, by assuming $V_s=exp(i\Phi_s)$, which is exponential of the phase operator, we obtain
 \begin{eqnarray}
\Phi_s =\frac{\pi s}{s+1}+ 2\pi\sum_{n\ne k}^s \frac{|n\rangle\langle k|}{e^{i\frac{2\pi}{s+1}[n-k]}-1},
\end{eqnarray}
where we have used Eq.(\ref{CF}) to obtain a closed form expression. This indicates that unitary transformations performed by circular waveguide arrays over discrete light fields will exhibit the same dynamics as the action of the phase operators over finite number states $N_{s}$. In the waveguide array each site represents a number state and the evolution operator is represented by the waveguide array itself. \\
Now we turn our attention to the field solution Eq.~(\ref{field1})
\begin{equation}
    \vec{E}(z)=\sum_{n=-\infty}^{\infty} J_n (-2\gamma z) i^nV_s^n  \vec{E}(0).
\end{equation}
The above equation may be rewritten as
\begin{equation}
    \vec{E}(z)=\sum_{n=0}^{\infty} J_n (-2\gamma z) i^nV_s^n  \vec{E}(0)+\sum_{n=1}^{\infty} J_{-n} (-2\gamma z) i^{-n}V_s^{\dagger n}\vec{E}(0),
\end{equation}
taking into account that $V^{s+1}=V^{\dagger (s+1)}=1$ we find that the sums become finite, for instance
\begin{equation}
    \sum_{n=0}^{\infty} J_n (-2\gamma z) i^nV_s^n =\sum_{n=0}^s F_nV_s^n
    \end{equation}
with $F_n=\sum_{k=0}^s i^{k(s+1)+n}J_{k(s+1)+n}(-2\gamma z)$. In addition, the property $V^{s+1}=V^{\dagger (s+1)}=1$ implies that self-imaging proccess can occur in these types of systems. As an example, in Fig.~(\ref{fig1}) we show the intensity evolution when light is injected into one the guides of a circular waveguide array having six channels.
\section{Conclusions}
By noting that position and momentum operators are related via the Fourier operator, we propose that the $(s+1)$-dimensional phase and number operators are related by the discrete Fourier transform, which naturally leads to the Pegg-Barnett phase operator. We have given a way to model functions of the Pegg-Barnett phase operator by propagating classical light in circular array waveguide arrays.

\end{document}